\documentclass[12pt]{article}

\usepackage{amssymb,amsmath,amsfonts,latexsym,graphicx}
\usepackage[numbers]{natbib}
\usepackage{hyperref}
\usepackage[justification=centering]{caption}
\usepackage{multirow}
\usepackage{subfig}

\begin{document}

\begin{center}
\Large \bf CASCRNet: An Atrous Spatial Pyramid Pooling and Shared Channel Residual based Network for Capsule Endoscopy \rm

\vspace{1cm}


\large K V Srinanda, \large  M Manvith Prabhu, \large Shyam Lal

\vspace{0.5cm}

\normalsize


Department of Electronics and Communication Engineering,\\ National Institute of Technology Karnataka, Surathkal - 575025, India

\vspace{5mm}


Email: {\tt \{srinandakv, prabhumanvith03, shyam.mtec\}@gmail.com}

\vspace{1cm}

\end{center}

\abstract{}
This manuscript summarizes work on the Capsule Vision Challenge 2024 by MISAHUB. To address the multi-class disease classification task, which is challenging due to the complexity and imbalance in the Capsule Vision challenge dataset, this paper proposes CASCRNet (Capsule endoscopy-Aspp-SCR-Network), a parameter-efficient and novel model that uses Shared Channel Residual (SCR) blocks and Atrous Spatial Pyramid Pooling (ASPP) blocks. Further, the performance of the proposed model is compared with other well-known approaches. The experimental results yield that proposed model provides better disease classification results. The proposed model was successful in classifying diseases with an F1 Score of 78.5\% and a Mean AUC of 98.3\%, which is promising given its compact architecture.

\section{Introduction}\label{sec1}
For a doctor, biomedical image classification requires large domain knowledge and expertise, and it takes even more for a machine to master it. With the advent of Deep Learning \cite{LeCun2015}, humans have been able to develop models capable of doing various tasks, of which biomedical image classification happens to be one. In \cite{PMID:34122785}, authors have discussed several machine learning and deep learning methods for biomedical classification on big datasets.

As proposed by authors of \cite{handa2024capsulevision2024challenge}, the challenge in hand invites contestants to create a model capable of automatically classifying abnormalities captured in video capsule endoscopy (VCE) video frames. The dataset \cite{Handa2024training} contains 37607 training images and 16132 validation images belonging to 10 classes each viz. Angioectasia, Bleeding, Erosion, Erythema, Foreign Body, Lymphangiectasia, Normal, Polyp, Ulcer and Worms. The testing set \cite{Handa2024testing} consists of 4385 images. The training and validation sets are highly imbalanced with Normal class being the large chunk of the data.

The authors of \cite{handa2024capsulevision2024challenge} have conducted a preliminary analysis utilizing a variety of machine-learning and deep-learning models such as ResNet50 \cite{DBLP:journals/corr/HeZRS15}, VGG16 \cite{simonyan2014very}, Support Vector Machines (SVM) \cite{708428} among others. DeepLab \cite{7913730} introduced atrous convolutions, also known as dilated convolutions, which increase the receptive field of the network without requiring additional parameters or downsampling, making them effective for dense prediction tasks like semantic segmentation. A key component of the proposed model is the incorporation of atrous convolution. RCCGNet \cite{Chanchal2023} utilized the concept of shared channel residual blocks for the classification of renal cell carcinoma and achieved good results while keeping the model compact. This study uses this approach in the proposed model with a few additions and modifications. This GitHub\footnote{\url{https://github.com/Manvith-Prabhu/CASCRNet.git}} repository contains the source code for this research. We were ranked \#2 on Validation set and \#15 on testing set performance.

\section{Methodology}\label{sec2}

\begin{figure}[h]
      \centering
      \includegraphics[width=120mm]{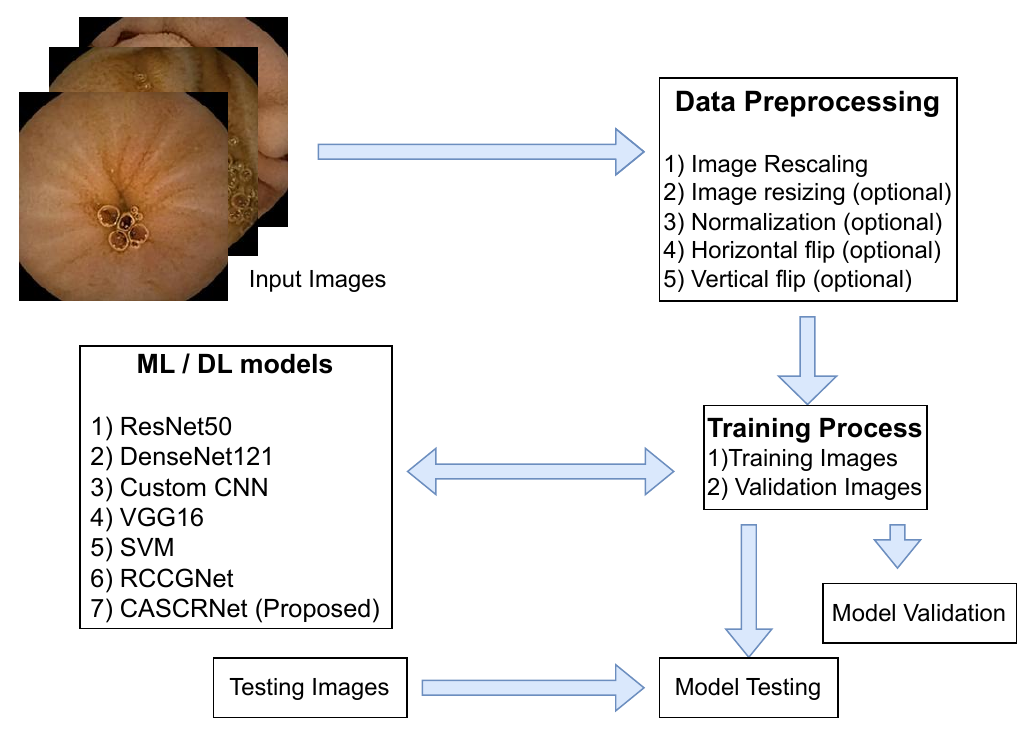}
      \caption{AI pipeline for Capsule Vision Challenge 2024}
      \label{fig:fig1}
\end{figure}

The challenge organizers experimented with various models, including ResNet50, VGG16, SVM, and a custom CNN, to establish a baseline. Among these, ResNet50, using a transfer learning approach, achieved the best performance when an input image size of 32x32 was used. Figure \ref{fig:fig1} shows the AI pipeline developed by our team for this challenge.

\subsection{ResNet50 and DenseNet121}
Models like ResNet50 \cite{DBLP:journals/corr/HeZRS15} and DenseNet121 \cite{8099726} have consistently demonstrated strong performance across various image classification tasks. For this challenge, these architectures with pre-trained weights from the ImageNet challenge were employed. By modifying the final layers with custom fully connected layers, the models were specifically fine-tuned for the Capsule Vision dataset, utilizing images with a resolution of 75x75 pixels. This method outperformed the organizers' approach by using larger image sizes and a fine-tuning strategy rather than relying on transfer learning. Also, the focal loss \cite{8237586} was utilized for training to deal with the highly imbalanced datasets. 

\subsection{RCCGNet}
\begin{figure}[h]
      \centering
      \includegraphics[width=65mm]{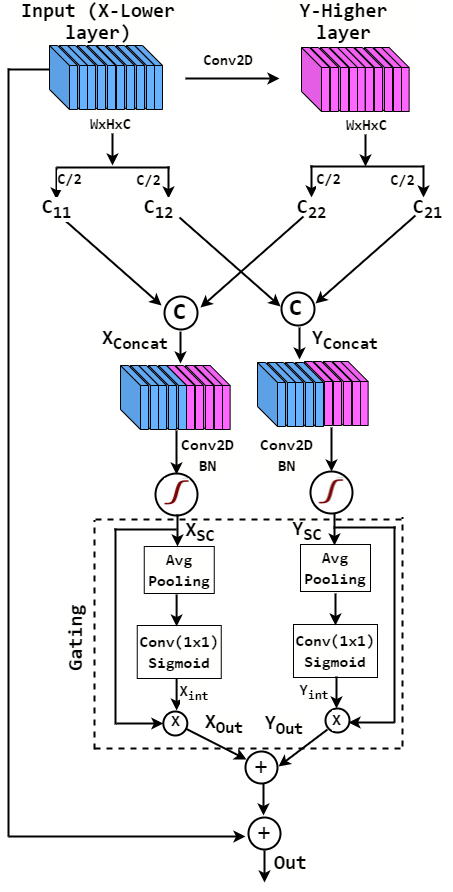}
      \caption{Shared Channel Residual block\\ Source: \cite{Chanchal2023}}
      \label{fig:fig2}
\end{figure}

\citeauthor{Chanchal2023} proposed RCCGNet model, a collaborative effort that included our co-author Dr. Shyam Lal. RCCGNet, as a lightweight model, which can be effectively trained with limited computational resources while accommodating full-sized input images of 224x224 pixels. The shared channel residual block in RCCGNet, illustrated in Figure \ref{fig:fig2}, concatenates two input layers before passing them through a pooling and convolution layer to generate the output. The RCCGNet architecture proposed in \cite{Chanchal2023} was implemented to achieve improved results. Focal loss was used to train RCCGNet because of its ability to deal with data imbalance.

\subsection{Proposed CASCRNet}
\begin{figure}[h]
      \centering
      \includegraphics[width=140mm]{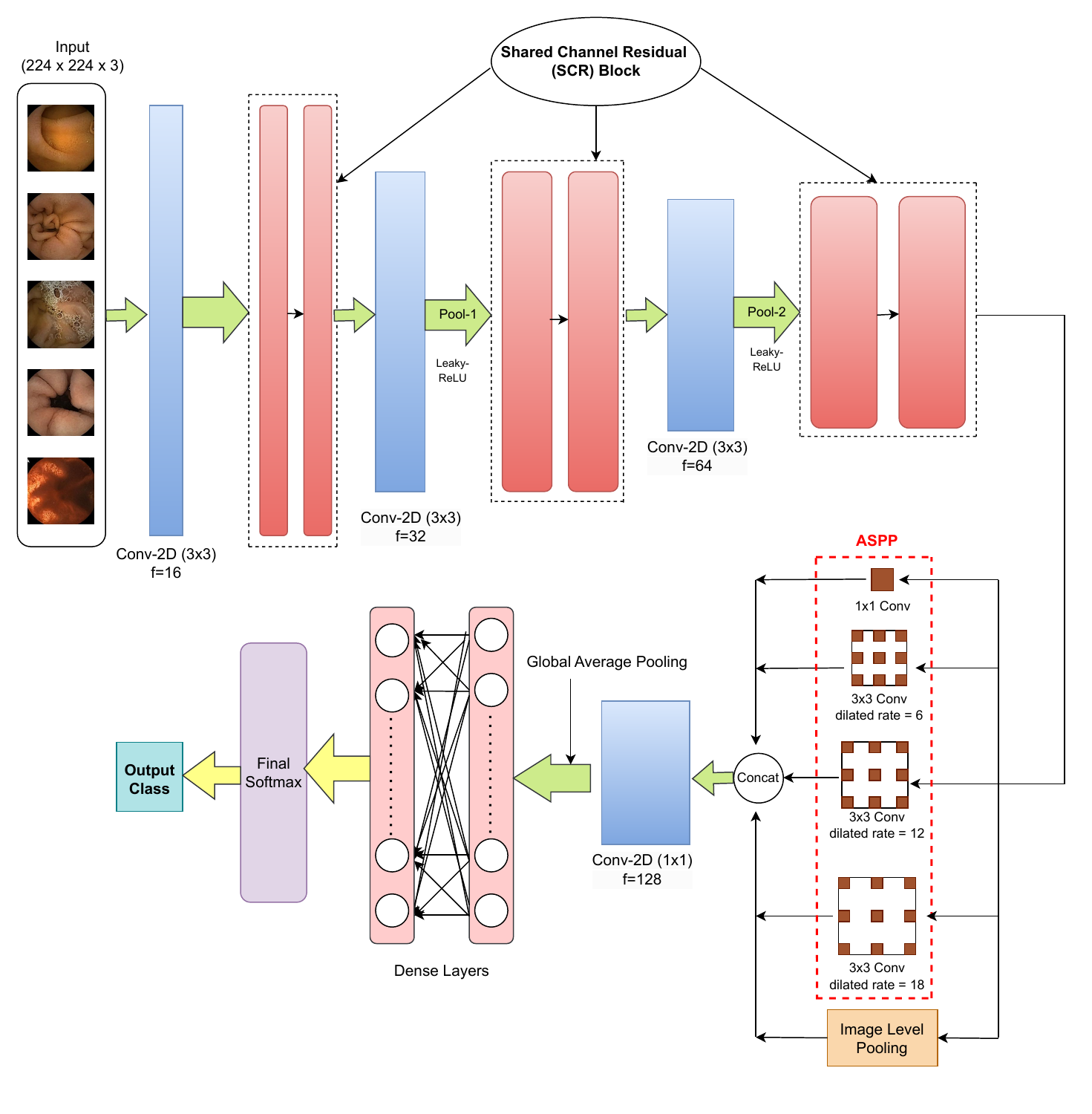}
      \caption{CASCRNet Architecture}
      \label{fig:fig3}
\end{figure}

The proposed model, CASCRNet, leverages the same shared channel residual block concept depicted in Figure \ref{fig:fig2}. An Atrous Spatial Pyramid Pooling block was incorporated to further enhance the model's ability to understand images. The LeakyReLU \cite{pmlr-v238-guo24c} activation function with an alpha value of 0.01 was utilized. In addition to this, the Adam optimizer \cite{kingma2014adam} was used, whose learning rate started with a value of 0.001 and was halved each time it reached a plateau. Dilated convolutions were applied within the residual blocks. Additionally, focal loss was implemented to tackle the class imbalance issue. The architecture of CASCRNet is illustrated in Figure \ref{fig:fig3}.

\section{Results}\label{sec3}

A comparison of the performance metrics of the submitted models and baseline models on the validation dataset is provided in Table \ref{tab:tab1}. The proposed model CASCRNet significantly outperformed the baselines created by the organizers of the challenge. The model exhibited strong performance, with a mean AUC of 98.3\% and a balanced accuracy of 77.7\%, both of which were key metrics emphasized in the challenge.

\begin{table*}[]
\centering
\begin{tabular}{|p{3cm}|p{1.3cm}|p{1.2cm}|p{1.3cm}|p{1.1cm}|p{1.3cm}|p{1.2cm}|p{1.2cm}|}
\hline
\textbf{Method} & \textbf{Image Size} & \textbf{\begin{tabular}[c]{@{}c@{}}Avg. \\ Acc.\end{tabular}} & \textbf{\begin{tabular}[c]{@{}c@{}}Avg. \\ Prec.\end{tabular}} & \textbf{\begin{tabular}[c]{@{}c@{}}Avg. \\ AUC\end{tabular}} & \textbf{\begin{tabular}[c]{@{}c@{}}Avg. \\ Recall\end{tabular}} & \textbf{\begin{tabular}[c]{@{}c@{}}Avg. \\ F1\end{tabular}} & \textbf{\begin{tabular}[c]{@{}c@{}}Bal. \\ Acc. \end{tabular}} \\ \hline
\textbf{SVM (baseline)} & 32x32 & 0.82 & 0.83 & 0.94 & 0.41  & 0.49 & 0.41\\ \hline
\textbf{ResNet50$\,^\dagger$ (baseline)} & 32x32 & 0.76 & 0.60 & 0.87 & 0.32  & 0.37 & 0.32 \\ \hline
\textbf{VGG16 (baseline)} & 32x32 & 0.69 & 0.52 & 0.92  & 0.54 & 0.48  & 0.57\\ \hline
\textbf{Custom CNN (baseline)} &224x224& 0.46 & 0.10 & 0.31 & 0.10  & 0.09 & 0.10\\ \hline
\textbf{ResNet50$\,^\S$} & 75x75 &0.912 & 0.818 & 0.978 & 0.749  & 0.784 & 0.759 \\ \hline
\textbf{DenseNet121} & 75x75 & 0.889 & 0.754 & 0.971 & 0.675  & 0.726 & 0.703\\ \hline
\textbf{RCCGNet} &224x224& 0.897 & 0.805 & 0.982 & 0.662  & 0.730 & 0.709\\ \hline
\textbf{CASCRNet (Proposed)} &224x224& \textbf{0.915} & \textbf{0.843} & \textbf{0.983} & \textbf{0.756}  & \textbf{0.785} & \textbf{0.777}\\ \hline
\end{tabular}
\caption{Comparison of the validation results to the baseline methods.\\ Transfer Learning$\,^\dagger$, Fine-tuning$\,^\S$}
\label{tab:tab1}
\end{table*}

\begin{figure}
    \centering
    \subfloat[ROC Curve]{\includegraphics[width=0.5\linewidth]{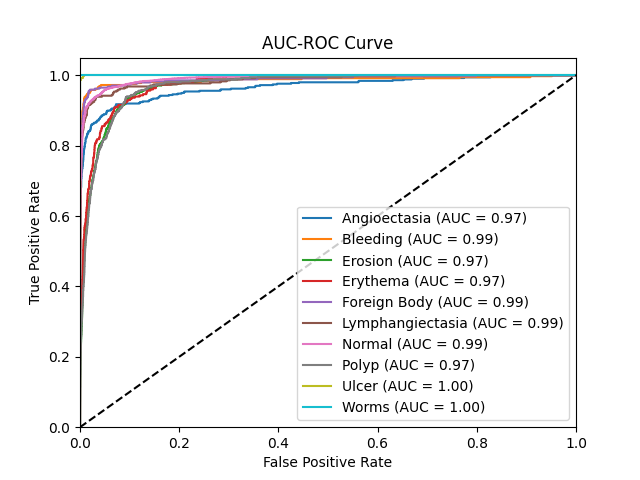}\label{fig:subfig4_a}}
    \hfill
    \subfloat[Normalised Confusion Matrix]{\includegraphics[width=0.5\linewidth]{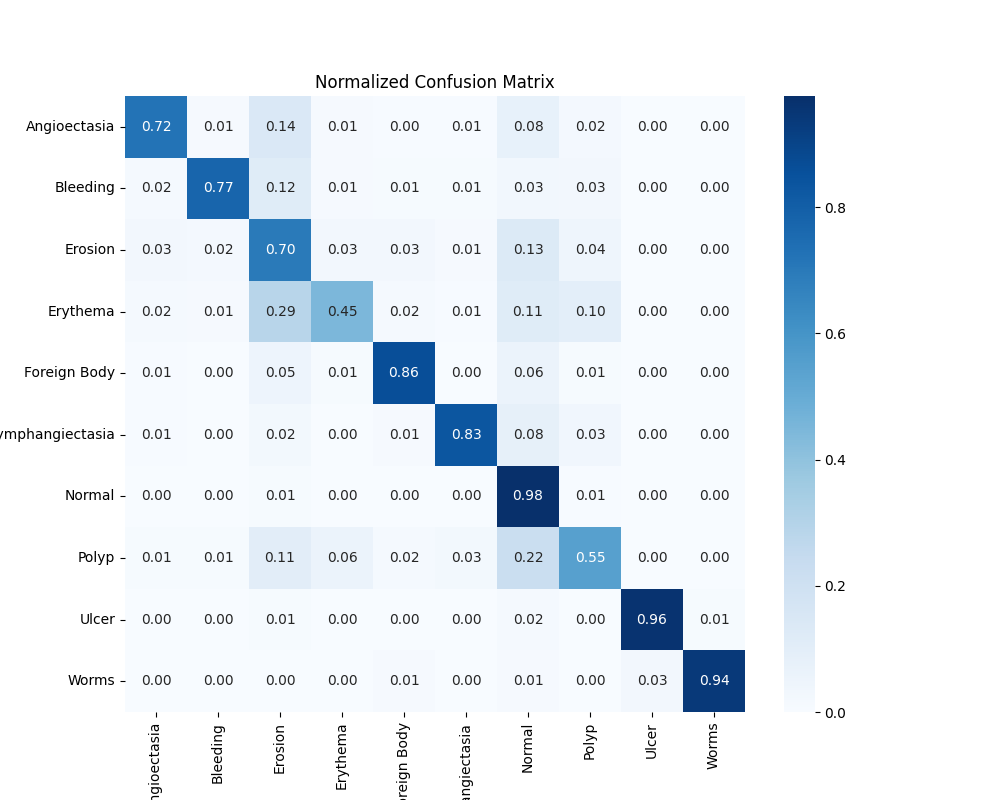}\label{fig:subfig4_b}}
    \caption{Performance of CASCRNet on the Validation Dataset}
    \label{fig:fig4}
\end{figure}

Figure \ref{fig:fig4} shows the Reciever Operating Characteristics (ROC) and normalized confusion matrix of CASCRNet on the validation set.

\section{Discussion}\label{sec4}

Table \ref{tab:tab1} clearly demonstrates that the proposed model, CASCRNet, outperforms all other models and baselines across every evaluation metric. Remarkably, CASCRNet achieves this while processing full-sized input images (224x224), even in the presence of a highly imbalanced training dataset. Despite its good performance, it is essential to note that the model is not intended as a replacement for a medical professional. 

\section{Conclusion and Future Scope}\label{sec5}

The proposed model, being both compact and parameter-efficient, is well-suited for deployment on edge devices and mobile platforms, making it an ideal tool to assist doctors and medical personnel. Moreover, the model can be extended for real-time monitoring in video capsule endoscopy, offering valuable clinical support. 

While the CASCRNet outperformed others in the comparative study, there remains room for improvement. Exploring architectures like Vision Transformers (ViTs) \cite{50650} and Vision-Language Models (VLMs) \cite{10445007} may lead to better-balanced accuracy. VLMs, in particular, could provide deeper insights and explain the rationale behind the model’s decisions, further enhancing its utility in medical applications.

\section{Acknowledgments}\label{sec6}
As participants in the Capsule Vision 2024 Challenge, we fully comply with the competition's rules as outlined in \cite{handa2024capsulevision2024challenge}. Our AI model development is based exclusively on the datasets provided in the official release in \cite{Handa2024training} and \cite{Handa2024testing}.

We would like to acknowledge Kaggle for providing the computational resources that supported this research. We also extend our thanks to the Department of Electronics and Communication at NITK Surathkal for offering additional computational resources, which were crucial for the successful completion of this work.

\bibliographystyle{unsrtnat}
\bibliography{main}

\end{document}